\documentclass
[prc,superscriptaddress,twocolumn,showpacs,preprintnu
mbers,amsmath,amssymb]{revtex4-1}
\usepackage[dvips]{graphicx}
\usepackage{dcolumn}
\usepackage{amsmath}
\usepackage{times}

\def\fun#1#2{\lower3.6pt\vbox{\baselineskip0pt\lineskip.9pt
 \ialign{$\mathsurround=0pt#1\hfil##\hfil$\crcr#2\crcr\sim\crcr}}}
\usepackage{graphicx}
\usepackage{dcolumn}
\usepackage{bm}

\newcommand{\beq}{\begin{equation}}
\newcommand{\eeq}{\end{equation}}
\newcommand{\bea}{\begin{eqnarray}}
\newcommand{\eea}{\end{eqnarray}}

\begin{document}


\title{Determination of $^8$B($p$,$\gamma$)$^9$C reaction rate from $^9$C breakup}


\author{Tokuro Fukui}
\email[Electronic address: ]{tokuro@rcnp.osaka-u.ac.jp}
\author{Kazuyuki Ogata}%
\affiliation{%
 Research Center for Nuclear Physics, Osaka University, Osaka 567-0047, Japan
}%

\author{Kosho Minomo}
\author{Masanobu Yahiro}
\affiliation{
 Department of Physics, Kyushu University, Fukuoka 812-8581, Japan
}%

\date{\today}

\begin{abstract}
The astrophysical factor of the
$^8$B($p$,$\gamma$)$^9$C at zero energy, $S_{18}(0)$, is determined
from three-body model analysis of $^9$C breakup processes.
The elastic breakup $^{208}$Pb($^9$C,$p{}^8$B)$^{208}$Pb at 65~MeV/nucleon and
the one-proton removal reaction of $^9$C at 285~MeV/nucleon on C and Al targets are
calculated with the continuum-discretized coupled-channels method (CDCC)
and the eikonal reaction theory (ERT), respectively.
The asymptotic normalization coefficient (ANC) of $^9$C in the
$p$-$^8$B configuration,
$C_{p {}^{8}\mbox{\scriptsize B}}^{^{9}\mbox{\scriptsize C}}$,
extracted from the two reactions show good consistency, in contrast to
in the previous studies. As a result of the present analysis,
$S_{18}(0)=66 \pm 10~{\rm eVb}$ is obtained.
\end{abstract}

\maketitle


{\it Introduction.}
In low-metallicity supermassive stars,
the proton capture reaction of $^8$B, $^8$B($p$,$\gamma$)$^9$C
ignites the explosive hydrogen burning~\cite{Wiescher}:
\[
^8{\rm B}(p,\gamma){}^9{\rm C}(\alpha,p){}^{12}{\rm N}(p,\gamma)
{}^{13}{\rm O}(\beta^+ \nu){}^{13}{\rm N}(p,\gamma){}^{14}{\rm O}.
\]
This process called hot pp chain is expected to be
a possible alternative path to the synthesis of the CNO elements.
Because of the difficulties in measuring the $^8$B($p$,$\gamma$)$^9$C
cross section $\sigma_{p\gamma}$
at very low energies, several alternative reactions have been
proposed~\cite{Beaumel,Trache,Motobayashi}
to indirectly determine
the astrophysical factor $S_{18}(\varepsilon)$
\beq
S_{18}(\varepsilon)=\sigma_{p\gamma} \varepsilon \exp[2\pi\eta].
\eeq
Here, $\varepsilon$ is the relative energy between $p$ and $^8$B in the
center-of-mass (c.m.) frame and $\eta$ is the Sommerfeld parameter.
Because an astrophysical factor has quite weak energy dependence,
several previous studies have paid special attention to the evaluation
of $S_{18}(\varepsilon)$ at zero energy,
$S_{18}(0)$~\cite{Wiescher,Beaumel,Trache,Motobayashi,Descouvemont}.

The Coulomb dissociation method~\cite{Motobayashi} is based on
the assumption
that elastic breakup of $^9$C by a heavy target, e.g., $^{208}$Pb,
is essentially
a one-step electric dipole (E1) transition to the $p+{}^8$B continuum.
Then $\sigma_{p\gamma}$ can be obtained by evaluating the cross section
of the inverse process of the breakup reaction~\cite{BB}.
This assumption needs to be examined, since nuclear breakup, Coulomb
dissociation with higher multipolarities, and multi-step transitions
can play non-negligible roles even in E1-dominated breakup
processes~\cite{S17}.
In fact, an attempt to evaluate these higher-order contributions
was made in Ref.~\cite{Motobayashi}; we will return to this point later.

The asymptotic normalization coefficient (ANC)
method~\cite{ANC}, which is one of the most important techniques
of indirect measurements have been used in several
studies~\cite{S17,Liu96,Oga03,Das06,Azh99,Tan04,TracheS17,sbat} in order
to determine astrophysical reaction rates.
The basic idea of the ANC method is
that only the tail of the overlap between the initial and final
states contributes to a reaction at stellar energies. Thus, the
purpose in the present case is to determine the ANC
$C_{p {}^{8}\mbox{\scriptsize B}}^{^{9}\mbox{\scriptsize C}}$
of the $^9$C wave function in the $p+{}^8$B configuration
by using some alternative reactions.
In Refs.~\cite{Beaumel} and \cite{Trache}, respectively,
the $d({}^8{\rm B},{}^9{\rm C})n$ reaction at 11.4~Mev/nucleon and
the one-proton removal reaction of $^9$C at 285~MeV/nucleon were
analyzed
to determine $C_{p {}^{8}\mbox{\scriptsize B}}^{^{9}\mbox{\scriptsize C}}$,
and hence $S_{18}(0)$.
One of the most important conditions for the ANC method is that
a reaction used to determine the ANC must be peripheral.
From this aspect, transfer reactions at low incident
energies~\cite{Liu96,Oga03,Das06,Azh99,Tan04,sbat}
and nucleon removal reactions in wide range of energies~\cite{TracheS17}
have been used as alternative reactions for the ANC method.
In Ref.~\cite{Cluster03}, it was demonstrated that an ANC can be
extracted from an elastic breakup cross section (angular distribution)
for which E1 breakup plays a dominant role. Later this method
was carefully examined and justified~\cite{S17}; important findings
of the work are i) E1-dominated breakup processes are
peripheral with respect to the relative coordinate between
the two fragments after the breakup,
ii) the breakup cross section in a coupled-channel framework
is proportional to
the square of the ANC to be determined, and iii) if the two fragments
are ejected in forward angles, which is the case in usual breakup
experiments of unstable nuclei, dynamical excitation of
each fragment during the breakup process has no essential effects
on the ANC.

\begin{table}[hptb]
\begin{center}
\caption{Astrophysical factors of $^8$B($p$,$\gamma$)$^9$C in previous studies.}
\begin{tabular}{|lcc|}
\hline
\hline
   &
  \multicolumn{1}{c}{$S_{18}$ [eVb]} &
  \multicolumn{1}{c|}{method}
\\
\hline
  \multicolumn{1}{|l}{Beaumel $et~al.~$\cite{Beaumel}} &
  \multicolumn{1}{c}{$45\pm13$} &
  \multicolumn{1}{c|}{ANC (transfer)}
\\
  \multicolumn{1}{|l}{Trache $et~al.~$\cite{Trache}} &
  \multicolumn{1}{c}{$46\pm6$} &
  \multicolumn{1}{c|}{ANC (proton removal)}
\\
  \multicolumn{1}{|l}{Motobayashi~\cite{Motobayashi}} &
  \multicolumn{1}{c}{$77\pm15$} &
  \multicolumn{1}{c|}{Coulomb dissociation}
\\
  \multicolumn{1}{|l}{Wiescher $et~al.~$\cite{Wiescher}} &
  \multicolumn{1}{c}{$210$} &
  \multicolumn{1}{c|}{shell model}
\\
  \multicolumn{1}{|l}{Descouvemont~\cite{Descouvemont}} &
  \multicolumn{1}{c}{$72,~80$} &
  \multicolumn{1}{c|}{cluster model}
\\
\hline
\hline
\end{tabular}
\label{previous}
\end{center}
\end{table}
We show in Table \ref{previous} the $S_{18}(0)$ reported in
the aforementioned indirect measurements~\cite{Beaumel,Trache,Motobayashi},
together with theoretical evaluations~\cite{Wiescher,Descouvemont}.
One sees that the two theoretical values have a large difference of about
factor of 3.
Experimental results seem to support the $S_{18}(0)$ obtained by a
cluster model calculation~\cite{Descouvemont}.
There is, however, still a significant discrepancy of about 50\% between the
$S_{18}(0)$ obtained by Coulomb dissociation method~\cite{Motobayashi} and
the ANC method~\cite{Beaumel,Trache}.

In this Rapid Communication, we reinvestigate the Coulomb dissociation~\cite{Motobayashi}
(elastic breakup)
and the proton removal process~\cite{Trache} of $^9$C by means of coupled-channel
calculation with a three-body ($p+{}^8{\rm B}+{\rm target}$) model.
We adopt the continuum-discretized
coupled-channels method (CDCC)~\cite{CDCC1,CDCC2,CDCC3}
for the former and the eikonal reaction theory (ERT)~\cite{ERT,ERT2} for the latter;
we use the ANC method for both reactions.
The main purpose of the present study is to show the consistency between
the two values of $S_{18}(0)$ extracted from these two types of breakup,
and thereby determine $S_{18}(0)$ with high reliability.

\begin{figure}[tb]
\begin{center}
\includegraphics[width=0.3\textwidth,clip]{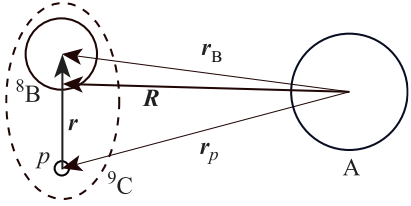}
\caption{Illustration of the three-body system.}
\label{9C-fig1}
\end{center}
\end{figure}

{\it Theoretical framework.}
In Fig.~\ref{9C-fig1} we show schematic illustration of the
three-body ($p+{}^8{\rm B}+{\rm target}$) system.
%
The scattering between $^9$C and a target nucleus A is
described by the Schr\"odinger equation
\beq
\left[
-\frac{\hbar^2}{2\mu}\nabla^2_{\bm{R}}
+h
+U(r_p,r_{\rm B})
-E
\right]
\Psi(\bm{r},\bm{R})=0,
\label{3bsc-eq}
\eeq
where $\Psi(\bm{r},\bm{R})$ is the tree-body wave function and
${\bm r}$ (${\bm R}$) is the coordinate of $^8$B ($^9$C) relative to $p$
(A). The reduced mass between $^9$C and A is denoted by $\mu$ and $E$
is the total energy of the three-body system in the c.m. frame.
The internal Hamiltonian of $^9$C is shown by $h$.
The interaction $U(r_p,r_{\rm B})$ is given by
\beq
U(r_p,r_{\rm B})
=V_p^{({\rm N})}(r_p)
+V_p^{({\rm C})}(r_p)
+V_{\rm B}^{({\rm N})}(r_{\rm B})
+V_{\rm B}^{({\rm C})}(r_{\rm B}),
\label{pot}
\eeq
where $V_{\rm X}^{({\rm N})}$ and $V_{\rm X}^{({\rm C})}$
are the nuclear and Coulomb interactions, respectively, between X and A;
X represents a fragment particle of the projectile, i.e., $p$ or $^8$B.
Similarly, $r_{\rm X}$ denotes the relative distance between
X and A.

In the present analysis of the elastic breakup of $^9$C, we solve
Eq.~(\ref{3bsc-eq}) with eikonal-CDCC (E-CDCC) \cite{S17,E-CDCC1}.
E-CDCC assumes eikonal approximation to the scattering wave between
$^9$C and A. As a result, the total wave function
$\Psi(\bm{r},\bm{R})$ is expressed by
\beq
\Psi({\bm r},{\bm R})
=
\sum_{c}
\Phi_{c}({\bm r})
e^{-i(m-m_0)\phi_R}
\psi_{c}(b,z)
\phi_{{\bm K}_c}^{{\rm C}}(b,z),
\label{wf-ecdcc}
\eeq
where $\Phi_{c}({\bm r})$ is the internal wave function of
$^9$C with $c$ the channel indices \{$i$, $\ell$, $S$, $I$, $m$\}; $i>0$
($i=0$) stands for the $i$th discretized-continuum (ground) state,
and $\ell$, $S$, and $I$ are, respectively, the orbital angular
momentum, the channel spin, and the total angular momentum
of the $p$ and $^8$B system. $m$ is the projection of $I$
on the $z$-axis taken to be parallel to the incident beam;
$m_0$ is the value of $m$ in the incident channel.
$b$ is the impact parameter defined by $b=\sqrt{x^2+y^2}$ with
${\bm R} = (x,y,z)$ in the Cartesian representation.
The use of the Coulomb incident wave
$\phi_{{\bm K}_c}^{{\rm C}}(b,z)$
instead of the plane wave $\exp ({\bm K}_c \cdot {\bm R})$
in the eikonal approximation is one of the most important
features of E-CDCC;
${\bm K}_c$ is the asymptotic wave-number vector of $^9$C in channel $c$
from A. In the actual calculation, we use an approximate asymptotic
form of $\phi_{{\bm K}_c}^{{\rm C}}(b,z)$.
E-CDCC is shown to work very well for describing both the nuclear
and Coulomb breakup processes with high accuracy and computational
speed~\cite{S17,E-CDCC1}.

The one-proton removal reaction, its
stripping component in fact (see below), is analyzed by means
of the eikonal reaction theory (ERT)~\cite{ERT,ERT2}, which
can calculate an inclusive cross section, such as a nucleon removal cross
section, in the CDCC framework.
ERT uses a formal solution (the scattering matrix $S$) to the
coupled-channel equations of E-CDCC, and makes adiabatic approximation
to only the nuclear part of $S$. Then one can obtain
the most important result of ERT, i.e., the product form of $S$~\cite{ERT}
\beq
S=S_{\rm b}S_{\rm c},
\eeq
where $S_{\rm b}$ and $S_{\rm c}$ show the contributions from
the constituents b and c of the projectile, respectively.
At this stage, however, this result can be derived only
when b or c is chargeless,
which is not the case for the $^9$C projectile consisting of
$p$ and $^8$B.
Therefore, in the present study, we neglect the Coulomb breakup
process in the one-proton removal process and replace the
Coulomb interaction $V_p^{({\rm C})}({\bm r}_p)$ with
\beq
V_p^{({\rm C})}(r_p) \rightarrow V_p^{({\rm C})}(R).
\label{replace}
\eeq
Then we can calculate the one-proton removal cross section
$\sigma_{-p}$ with
\beq
\sigma_{-p} = \sigma_{\rm bu} + \sigma_{\rm str},
\label{removalcs}
\eeq
as in Refs.~\cite{ERT,ERT2}.
In Eq.~(\ref{removalcs}), $\sigma_{\rm bu}$ and $\sigma_{\rm str}$
denote the elastic breakup cross section and the stripping
cross section, respectively; ERT is used to evaluate $\sigma_{\rm str}$.
The accuracy of the replacement of Eq. (\ref{replace})
can be examined by calculating $\sigma_{-p}$ with
and without the Coulomb breakup. It is confirmed that
the Coulomb breakup contributes to $\sigma_{-p}$
for C and Al targets by about 5\%.
Thus, we conclude that the Coulomb breakup
by these two targets can be neglected with 5\% errors.
Below we include this amount in uncertainties of $S_{18}(0)$
extracted from $\sigma_{-p}$.

{\it Model setting.}
For both the elastic breakup and one-proton removal processes,
the $p$-$^8$B wave function is calculated with
the same Hamiltonian $h$. We include only the intrinsic spin of $p$.
We adopt the standard Woods-Saxon central potential with the radial
parameter $R_0 = 1.25\times 8^{1/3}~{\rm fm}$ and
the diffuseness parameter $a_0 = 0.65~{\rm fm}$.
The Coulomb interaction between a point charge ($p$)
and a uniformly charged sphere ($^8$B) with the charge radius
of 2.5~fm is included.
For the p-wave states, we add the Thomas-type spin-orbit
interaction, with the same $R_0$ and $a_0$ as of the central part.
The depth of the spin-orbit is set to 4.40~MeV, and that of the
central part is determined to reproduce the proton separation
energy $S_p=1.30$~MeV in the $3/2^-$ state.
With this potential, we have a resonance
state at $\varepsilon=0.915$~MeV with the width $\Gamma=0.137$~MeV
in the $1/2^-$ state, in good agreement with the experimental values,
i.e., $\varepsilon=0.918\pm0.011$~MeV and
$\Gamma=100\pm 20$~keV~\cite{exp9C}.
We include s1/2$^+$, p1/2$^-$, p3/2$^-$, d3/2$^+$, d5/2$^+$, f5/2$^-$,
and f7/2$^-$ waves of the $p+{}^8$B system in the coupled-channel
calculations.

As for the nuclear part of the distorting potential
$V^{\rm (N)}_{\rm X} ({\rm X}=p~{\rm or}~{}^8{\rm B})$,
we adopt the microscopic folding model~\cite{fold1,fold2}
with the Melbourne nucleon-nucleon $g$ matrix~\cite{Mel}. Nuclear
densities of $^8$B, $^{12}$C,  $^{27}$Al, and $^{208}$Pb
are calculated by Hartree-Fock (HF) method with
the Gogny-D1S force~\cite{Gogny1,Gogny2}.
The resulting microscopic proton optical potentials are found to
reproduce, with no adjustable parameters,
the elastic scattering cross sections for
$p$-$^{208}$Pb at 65~MeV~\cite{Sakaguchi}
and the $p$-$^{12}$C reaction cross sections at
200--400~MeV~\cite{NNDC}.
For $^8$B-A scattering, however, it turns out that a fine tuning
of the optical potential is necessary. This can be done with
replacing the argument of both the real and imaginary parts of
$V^{\rm (N)}_{\rm X}$ as
\beq
r_{\rm B} \rightarrow (1+x) r_{\rm B},
\label{widen}
\eeq
which effectively increase the range of the potential.
We set $x$ to 0.04 (0.03) for the $^8$B-$^{12}$C ($^8$B-$^{27}$Al)
potential at 285~MeV/nucleon to reproduce the experimental data of the
reaction cross section~\cite{1p-exp}.
As for the $^8$B-$^{208}$Pb reaction at 65~MeV/nucleon, since there is
no experimental data, we calculate the reaction cross section
by CDCC with a $p+{}^7{\rm Be}+{}^{208}$Pb three-body model,
and $x=0.10$ is obtained to reproduce the calculated value.
The prescription of Eq.~(\ref{widen}) can be understood as
a modification of the HF density of $^8$B to include a halo
structure effectively.

\begin{figure}[hptb]
\begin{center}
\includegraphics[width=0.4\textwidth,clip]{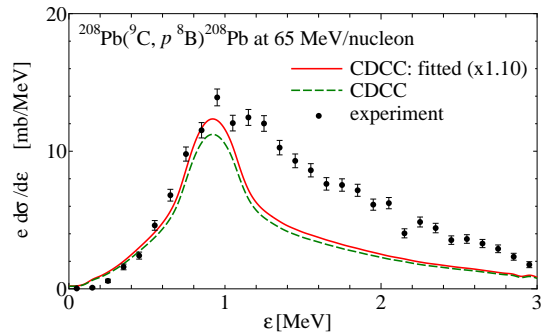}
\caption{(Color online)
 Breakup spectrum of the $^{208}$Pb($^9$C,$p{}^8$B)$^{208}$Pb at
 65~MeV/nucleon as a function of the relative energy $\varepsilon$ between $p$
 and $^8$B. The dashed line shows the result of calculation with a normalized
 $p$-$^8$B wave function, whereas the solid line is the result multiplied by
 1.1 to fit the experimental data~\cite{Motobayashi}.
}
\label{9C-fig2}
\end{center}
\end{figure}

\begin{table}[hptb]
\begin{center}
\caption{Model space of the present calculation.
See the text for details.}
\begin{tabular}{|l|c|c|}
\hline
\hline
  Reaction &
  elastic breakup &
  proton removal
\\
\hline
$k_{\rm max}~{\rm [fm^{-1}]}$ &
  1.0 &
  1.2
\\
$\Delta k~{\rm [fm^{-1}]}$ &
  0.05 &
  0.10
\\
$r_{\rm max}~{\rm [fm]}$ &
  150 &
  150
\\
$R_{\rm max}~{\rm [fm]}$ &
  250 &
  30
\\
$L_{\rm max}$ &
  2,000 &
  450
\\
\hline
\hline
\end{tabular}
\label{modelspace}
\end{center}
\end{table}
The model space of the present CDCC calculation is summarized in
Table~\ref{modelspace}, where $k_{\rm max}$ ($r_{\rm max}$) is
the maximum value of the relative wave number $k$ (coordinate $r$)
between $p$ and $^8$B, and $\Delta k$ represents the width
of the momentum bin.
$R_{\rm max}$ and $L_{\rm max}$ are, respectively, the maximum
values of the relative
coordinate and the orbital angular momentum between $^9$C and A.
We have confirmed with the model space the convergence of the
elastic breakup cross section (Fig.~\ref{9C-fig2})
for $\varepsilon \le 1$~MeV and $\sigma_{-p}$ (Table \ref{removal_cs})
both within 1\%.

{\it Results and discussion.}
First, we analyze the elastic breakup $^{208}$Pb($^9$C,$p{}^8$B)$^{208}$Pb
at 65~MeV/nucleon.
%
In Fig. \ref{9C-fig2}, we show the breakup cross section
as a function of the relative energy $\varepsilon$ between $p$
and $^8$B.
We have included the experimental efficiency $e(\varepsilon)$~\cite{private}
and resolution $\Gamma$ in the calculation.
We adopt $\Gamma = 0.23$~MeV extracted from the
experimental breakup spectrum of $^{12}$C($^9$C,$p{}^8$B)$^{12}$C
at 65~MeV/nucleon~\cite{private}.
In order to determine
$C_{p {}^{8}\mbox{\scriptsize B}}^{^{9}\mbox{\scriptsize C}}$,
we fit the theoretical result (dashed line)
to the experimental data~\cite{Motobayashi},
and the solid line is obtained. The renormalization factor is 1.10,
which results in
$(C_{p {}^{8}\mbox{\scriptsize B}}^{^{9}\mbox{\scriptsize
C}})^2=1.78$~fm$^{-1}$ and $S_{18}(0)=67.3$~eVb.

In Fig. \ref{9C-fig2}, our calculation describes well the
breakup spectrum below $\varepsilon \sim 1.0$~MeV, i.e., both the
transition to the $1/2^-$ resonant state and breakup to
low-energy nonresonant states of $^9$C.
It should be noted that we treat the
resonant and nonresonant breakup continua on the same footing
in the CDCC calculation.
In the higher $\varepsilon$ region than the resonance energy, however,
the calculation significantly underestimates the experimental data.
It is expected that this is due to incompleteness of our
present framework. The back-coupling effects of
three-body breakup states of $^9$C to $p+p+{}^7$Be on the
$p+{}^8$B state observed will become important as $\varepsilon$
increases. In addition, more accurate description of the $p+{}^8$B
continua for higher partial waves with a proper $p$-$^8$B
interaction $V_{p\rm{B}}^{\rm (N)}$ will be needed. At low $\varepsilon$,
these possible problems will not exist, because only the tail
of the overlap between $^9$C and $p$-$^8$B contributes to the
breakup process.
For more detailed discussion on this point, see Ref.~\cite{S17}.

To examine the peripherality of the $^{208}$Pb($^9$C,$p{}^8$B)$^{208}$Pb
reaction, we see the dependence of
$C_{p {}^{8}\mbox{\scriptsize B}}^{^{9}\mbox{\scriptsize C}}$
on the parameters of $V_{p\rm{B}}^{\rm (N)}$; each of $R_0$ and
$a_0$ is changed  by 20\%.
Note that we put a constraint on the depth of the central potential
so that it must reproduce the proton separation energy $S_p$.
It is found that the uncertainty of
$C_{p {}^{8}\mbox{\scriptsize B}}^{^{9}\mbox{\scriptsize C}}$
regarding $V_{p\rm{B}}^{\rm (N)}$ is 8\%.
This indicates that the present elastic breakup reaction proceeds
peripherally with respect to ${\boldsymbol r}$, as required by
the ANC method.

\begin{table}[hptb]
\begin{center}
\caption{Results of the one-proton removal reactions with
$^{12}$C and $^{27}$Al targets.
The experimental data of $\sigma_{-p}$ are taken from
Ref.~\cite{1p-exp}.
}
\begin{tabular}{|l|cc|cc|}
\hline
\hline
  Target &
  \multicolumn{2}{c|}{$^{12}$C} &
  \multicolumn{2}{c|}{$^{27}$Al}
\\
\hline
 &
  \multicolumn{1}{c}{\; calc. \;} &
  \multicolumn{1}{c|}{\; expt. \;} &
  \multicolumn{1}{c}{\; calc. \;} &
  \multicolumn{1}{c|}{\; expt. \;}
\\
\hline
\; $\sigma_{\rm bu}~{\rm [mb]}$ &
  \multicolumn{1}{r}{}2.7 &
  \multicolumn{1}{r|}{} &
  \multicolumn{1}{r}{4.7} &
  \multicolumn{1}{r|}{}
\\
\; $\sigma_{\rm str}~{\rm [mb]}$ &
  \multicolumn{1}{r}{42.2} &
  \multicolumn{1}{r|}{} &
  \multicolumn{1}{r}{49.2} &
  \multicolumn{1}{r|}{}
\\
\; $\sigma_{-p}~{\rm [mb]}$ &
  \multicolumn{1}{r}{44.9} &
  \multicolumn{1}{r|}{48(8)} &
  \multicolumn{1}{r}{53.9} &
  \multicolumn{1}{r|}{55(11)}
\\
\hline
\; $(C_{p {}^{8}\mbox{\scriptsize B}}^{^{9}\mbox{\scriptsize C}})^2
~~{\rm [fm^{-1}]}$ &
  \multicolumn{1}{r}{1.73} &
  \multicolumn{1}{r|}{} &
  \multicolumn{1}{r}{1.65} &
  \multicolumn{1}{r|}{}
\\
\; $S_{\rm 18}(0)~{\rm [eVb]}$ &
  \multicolumn{1}{r}{65.2} &
  \multicolumn{1}{r|}{} &
  \multicolumn{1}{r}{62.2} &
  \multicolumn{1}{r|}{}
\\
\hline
\hline
\end{tabular}
\label{removal_cs}
\end{center}
\end{table}
Second, we analyze the one-proton removal reaction of $^9$C
at 285~MeV/nucleon on $^{12}$C and $^{27}$Al targets.
As already mentioned, we neglect the Coulomb breakup of $^9$C in this case.
We calculate $\sigma_{\rm bu}$ by CDCC and the
stripping cross section $\sigma_{\rm str}$ by ERT, and obtain
the one-proton removal cross section $\sigma_{-p}$, as the
sum of the two.
Then we renormalize the calculated $\sigma_{-p}$ to fit the experimental
value taken from Ref.~\cite{1p-exp}, which determines
$(C_{p {}^{8}\mbox{\scriptsize B}}^{^{9}\mbox{\scriptsize C}})^2$
and hence $S_{\rm 18}(0)$.
These values are summarized in Table~\ref{removal_cs}.
One sees that the two results of $S_{\rm 18}(0)$,
corresponding to $^{12}$C and $^{27}$Al targets,
agree well with each other.
By taking an average of the two values,
we obtain
$(C_{p {}^{8}\mbox{\scriptsize B}}^{^{9}\mbox{\scriptsize C}})^2 = 1.69$~fm$^{-1}$
and $S_{18}(0)=63.7$~eVb.
In order to evaluate the uncertainty of the ANC for the one-proton removal
reactions, we take the same procedure as in the analysis of
the elastic breakup reaction; the uncertainty turns out to be 20\%.
By adding the aforementioned 5\% uncertainty due to the neglect of
Coulomb breakup, we find the total uncertainty of $S_{18}(0)$
extracted from $\sigma_{-p}$ to be 21\%.

We here remark that in our three-body coupled-channel analysis,
the values of $S_{18}(0)$ extracted from two different breakup reactions,
67.3~eVb (elastic breakup) and 63.7~eVb (proton removal),
show very good agreement. This indicates reliability
of the present analysis and the result of $S_{18}(0)$.
As a principal result of the present study, we obtain
$(C_{p {}^{8}\mbox{\scriptsize B}}^{^{9}\mbox{\scriptsize C}})^2
=1.7 \pm 0.3$~fm$^{-1}$, which corresponds to
\beq
S_{18}(0)
=66 \pm 10~{\rm eVb}.
\eeq
%
\begin{figure}[hptb]
\begin{center}
\includegraphics[width=0.4\textwidth,clip]{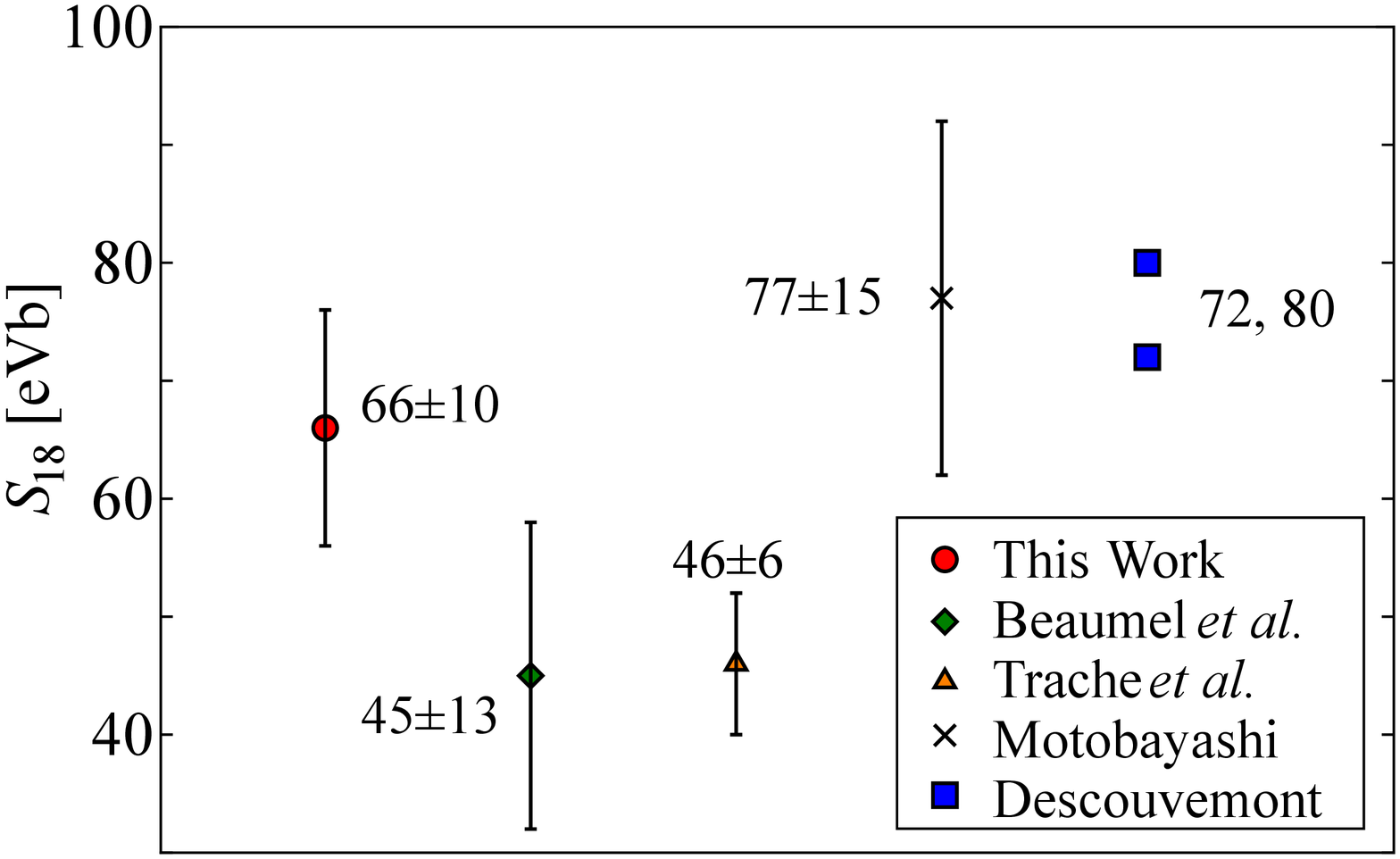}
\caption{
(Color online)
$S_{18}(0)$ extracted by this work (circle) is compared with
the results of the Coulomb dissociation method (cross)~\cite{Motobayashi}
and the analysis of $\sigma_{-p}$ with the extended Glauber model
(triangle)~\cite{Trache}.
Theoretical results with a cluster model calculation
(squares)~\cite{Descouvemont}
and the value extracted from
the $d({}^8{\rm B},{}^9{\rm C})n$ reaction (diamond)~\cite{Beaumel}
are also shown.
}
\label{9C-fig3}
\end{center}
\end{figure}
In Fig. \ref{9C-fig3}, the $S_{18}(0)$ extracted by the present work
is compared with previous values.
As mentioned above, previous results can be categorized into
two, i.e., one is around 80~eVb (Ref.~\cite{Descouvemont,Motobayashi})
and the other is around 45~eVb (Ref.~\cite{Beaumel,Trache}).
Our result exists in between them, slightly
favoring the former.

In Ref.~\cite{Motobayashi}, the E1 contribution to the
elastic breakup of $^9$C by $^{208}$Pb at
65~MeV/nucleon was extracted by subtracting the
contributions of the nuclear and E2 breakup processes ($\sim 10$\%) from
the measured cross section, with a help of the $^9$C breakup data
by $^{12}$C at the same energy.
The rather good consistency between the present and previous results
of $S_{18}(0)$
will indicate that the procedure for extracting the E1 contribution
worked quite well.
It was reported in Ref.~\cite{Motobayashi}, however, that
about 80\% of the peak in the $^{208}$Pb($^9$C,$p{}^8$B)$^{208}$Pb
breakup spectrum around $\varepsilon=0.9$~MeV was explained by nonresonant
E1 breakup processes. On the other hand,
in the present analysis, the peak is found to be
mainly generated by the nuclear and E2 transition
to the $1/2^-$ resonance state.
Reason for this large discrepancy in the resonant part
between the present and previous studies
needs further investigation; this is our important future work. 
If we adopt a one-step calculation including nuclear and Coulomb
breakup with all multipolarities,
$S_{18}(0)=54$~eVb is obtained, i.e., 20\% difference appears.
This behavior is the same as in the study of $S_{17}(0)$ for the
$^7$Be($p$,$\gamma$)$^8$B reaction~\cite{S17}.

Our result is quite larger than the result of Ref.~\cite{Trache},
in which the one-proton removal reactions ($^9$C,$^8$B)
at 285~MeV/nucleon were analyzed by the extended Glauber model,
with carefully evaluating the uncertainty regarding
the nucleon-nucleon effective interactions (profile functions).
By a detailed analysis, it is found that the difference
between the $S_{18}(0)$ obtained in the present work and
Ref.~\cite{Trache} is mainly due to the proton optical potential.
In Fig.~4 of Ref.~\cite{Trache}, the reaction cross section
$\sigma_{\rm R}$
of the $p$-$^{12}$C (solid line) is compared with experimental
data. As shown in the figure, the data have quite large
uncertainty; there seem to be two data groups between 250~MeV
and 600~MeV. Our microscopic calculation based on the
Melbourne $g$ matrix gives $\sigma_{\rm R}=198$~mb
at 285~MeV, which is smaller than the value
used in the previous study by about 10\%.
It should be noted that both the theoretical values of $\sigma_{\rm R}$
are consistent with the experimental data, within their uncertainty
mentioned above.
This 10\% difference is indeed crucial
for the evaluation of $\sigma_{-p}$, which eventually gives
the difference in $S_{18}(0)$ by about 35\%.
Thus, more accurate and reliable data of $\sigma_{\rm R}$
is highly desirable to judge the microscopic theoretical
calculations of $\sigma_{\rm R}$,
although we have shown in this study a very good agreement
between the two $S_{18}(0)$ extracted from different breakup
reactions.

Very recently, ANCs for light nuclei with mass numbers between
3 and 9 are systematically evaluated by a variational
Monte Carlo calculation~\cite{NW11}. The resulting value of
$(C_{p {}^{8}\mbox{\scriptsize B}}^{^{9}\mbox{\scriptsize C}})^2$
to be compared with ours ($1.7 \pm 0.3$~fm$^{-1}$) reads
$1.36 \pm 0.03$~fm$^{-1}$. It will be interesting to investigate
the difference between the two values in more detail.

{\it Summary.}
We have analyzed
the elastic breakup of $^9$C by $^{208}$Pb at 65~MeV/nucleon and
the one-proton removal reaction of $^9$C at 285~MeV/nucleon on C and Al targets
by a three-body coupled-channel framework, i.e., CDCC for the elastic
breakup process and ERT for the stripping process.
We determined the ANC
$C_{p {}^{8}\mbox{\scriptsize B}}^{^{9}\mbox{\scriptsize C}}$
and obtained the astrophysical factor at zero energy, $S_{18}(0)$,
for the $^8$B($p$,$\gamma$)$^9$C reaction.
Our principal result is $S_{18}(0)=66 \pm 10$~eVb.
We have confirmed that the results of $S_{18}(0)$ extracted from the two independent
experimental data agree very well with each other, and thus resolved
a significant discrepancy of $S_{18}(0)$ in the previous studies.
Although the ANC is determined well in the present analysis,
description of the breakup spectrum at higher $p$-$^8$B
relative energies is not sufficient. Extension of the present
reaction model to incorporate the $p+p+{}^7$Be configuration
will be very important for deeper understanding of the breakup
of $^9$C. Investigation on the $d({}^8{\rm B},{}^9{\rm C})n$
transfer reaction, which gives a quite smaller $S_{18}(0)$ than
in the present study, will also be important.

\bigskip
The authors wish to thank T. Motobayashi and Y. Togano for helpful discussions and
 providing experimental information on the elastic breakup reaction.
The computation was carried out using the computer facilities at the
 Research Institute for Information Technology, Kyushu University.
This research was supported in part by Grant-in-Aid of the Japan
Society for the Promotion of Science (JSPS).

\nocite{*}



\end{document}